\documentclass{emulateapj}
\usepackage{apjfonts}
\usepackage{rotating}

\newcommand{\pivec}{\mbox{\boldmath $\pi$}}

\lefthead{SHIN ET AL.} 
\righthead{ORBITAL-BASED RESOLUTION OF CLOSE/WIDE BINARY DEGENERACY
}

\begin{document}
\title{
Using Orbital Effects to Break the Close/wide Degeneracy in 
Binary-lens Microlensing Events
}

\author{
I.-G. Shin$^{001}$,
T. Sumi$^{101,100}$,
A. Udalski$^{201,200}$,    
J. Y. Choi$^{001}$,
C. Han$^{001,900}$,
A. Gould$^{301}$\\ 
and\\
F.\ Abe$^{102}$,        
D.\ P.\ Bennett$^{103}$,  
I.\ A.\ Bond$^{104}$,
C.\ S.\ Botzler$^{105}$,
P.\ Chote$^{106}$,
M.\ Freeman$^{105}$,   
A.\ Fukui$^{107}$,   
K.\ Furusawa$^{102}$,
P.\ Harris$^{106}$,   
Y.\ Itow$^{102}$,        
C.\ H.\ Ling$^{104}$,     
K.\ Masuda$^{102}$,      
Y.\ Matsubara$^{102}$,    
N.\ Miyake$^{102}$,
Y.\ Muraki$^{108}$,     
K.\ Ohnishi$^{109}$,    
N.\ Rattenbury$^{105}$,   
To.\ Saito$^{110}$,    
D.\ J.\ Sullivan$^{106}$,  
D.\ Suzuki$^{101}$,      
W.\ L.\ Sweatman$^{104}$, 
P.\ J.\ Tristram$^{111}$,  
K.\ Wada$^{101}$,        
P.\ C.\ M. Yock$^{105}$\\   
(The MOA Collaboration),\\
M.\ K.\ Szyma{\'n}ski$^{201}$,
M.\ Kubiak$^{201}$, 
I.\ Soszy{\'n}ski$^{201}$,
G.\ Pietrzy{\'n}ski$^{201,202}$, 
R.\ Poleski$^{201}$, 
K.\ Ulaczyk$^{201}$, 
P.\ Pietrukowicz$^{201}$,
S.\ Koz{\l}owski$^{201}$,
J.\ Skowron$^{301}$,
and
{\L}. Wyrzykowski$^{201,203}$\\ 
(The OGLE Collaboration)\\
%
% ----------------------------------
}

\bigskip\bigskip
% -------------------------------------------------------------------------------------------------------------------------------------------------------------
\affil{$^{001}$Department of Physics, Institute for Astrophysics, Chungbuk National University, Cheongju 371-763, Korea}
\affil{$^{101}$Department of Earth and Space Science, Osaka University, Osaka 560-0043, Japan}
\affil{$^{201}$Warsaw University Observatory, Al. Ujazdowskie 4, 00-478 Warszawa, Poland}
\affil{$^{301}$Department of Astronomy, Ohio State University, 140 W. 18th Ave., Columbus, OH 43210, USA}
% MOA ---------------------------------------------------------------------------------------------------------------------------------------------------------
\affil{$^{102}$Solar-Terrestrial Environment Laboratory, Nagoya University, Nagoya, 464-8601, Japan}
\affil{$^{103}$Department of Physics, University of Notre Damey, Notre Dame, IN 46556, USA}
\affil{$^{104}$Institute of Information and Mathematical Sciences, Massey University, Private Bag 102-904, North Shore Mail Centre, Auckland, New Zealand} 
\affil{$^{105}$Department of Physics, University of Auckland, Private Bag 92019, Auckland, New Zealand}
\affil{$^{106}$School of Chemical and Physical Sciences, Victoria University, Wellington, New Zealand} 
\affil{$^{107}$Okayama Astrophysical Observatory, National Astronomical Observatory, 3037-5 Honjo, Kamogata, Asakuchi, Okayama 719-0232, Japan}  
\affil{$^{108}$Department of Physics, Konan University, Nishiokamoto 8-9-1, Kobe 658-8501, Japan} 
\affil{$^{109}$Nagano National College of Technology, Nagano 381-8550, Japan} 
\affil{$^{110}$Tokyo Metropolitan College of Industrial Technology, Tokyo 116-8523, Japan}
\affil{$^{111}$Mt. John Observatory, P.O. Box 56, Lake Tekapo 8770, New Zealand} 
% OGLE --------------------------------------------------------------------------------------------------------------------------------------------------------
\affil{$^{202}$Universidad de Concepci\'{o}n, Departamento de Astronomia, Casilla 160-C, Concepci\'{o}n, Chile}
\affil{$^{203}$Institute of Astronomy, University of Cambridge, Madingley Road, Cambridge CB3 0HA, United Kingdom}
% --------------------------------------------------------------------------------------------------------------------------
\affil{$^{100}$The MOA Collaboration}
\affil{$^{200}$The OGLE Collaboration}
\affil{$^{900}$Corresponding author} 
% --------------------------------------------------------------------------------------------------------------------------

\begin{abstract}
Microlensing can provide an important tool to study binaries, especially 
those composed of faint or dark objects. However, accurate analysis of 
binary-lens light curves is often hampered by the well-known degeneracy 
between close $(s<1)$ and wide $(s>1)$ binaries, which can be very 
severe due to an intrinsic symmetry in the lens equation. Here $s$ is 
the normalized projected binary separation. In this paper, we propose 
a method that can resolve the close/wide degeneracy using the effect of 
a lens orbital motion on lensing light curves. The method is based on 
the fact that the orbital effect tends to be important for close binaries 
while it is negligible for wide binaries. We demonstrate the usefulness 
of the method by applying it to an actually observed binary-lens event 
MOA-2011-BLG-040/OGLE-2011-BLG-0001, which suffers from severe close/wide 
degeneracy. From this, we are able to uniquely specify that the lens is 
composed of K and M-type dwarfs located at $\sim$3.5 kpc from the Earth.
\end{abstract}


\section{Introduction}

Currently, about 2000 microlensing events are being detected each 
year from survey experiments (OGLE: Udalski 2003; MOA: Bond et al. 2001; 
Sumi et al. 2003). Among them, a considerable fraction are produced by 
objects composed of two masses. When a lensing event is produced by a 
binary object, the resulting light curve deviates from the smooth and 
symmetric form of a single-lens event \citep{mao91}. Analyzing a 
binary-lens light curve is important because it enables to extract 
information about the lens. For general binary-lens events, this 
information includes the binary mass ratio, $q$, and the projected 
separation between the lens components normalized by the angular 
Einstein radius $\theta_{\rm E}$ of the lens, $s$ (normalized projected 
separation). For events with resolved caustic crossings, it is possible 
to measure or constrain the mass and distance to the lens. Not being 
dependent on the lens luminosity, microlensing provides an important 
tool especially for the study of binaries composed of faint or dark objects.

However, accurate analysis of a binary-lens light curve is often 
hampered by the ambiguity in fits of observed light curves. Many 
cases of degenerate solutions are accidental, implying that the 
degeneracy can be resolved by precise and dense coverage of light 
curves. However, due to a symmetry in the lens equation itself, 
degeneracies between close $(s<1)$ and wide $(s>1)$ binaries can be 
very severe. This was first pointed out for a special case of 
low-mass (planetary) companions by \citet{griest98}.  It was then 
generalized to all binaries by \citet{dominik99} and analyzed still 
more thoroughly by \citet{an05}. Therefore, devising a method resolving
the close/wide degeneracy is important to uniquely characterize binary 
lens systems.

In this paper, we propose a method that can resolve the close/wide 
degeneracy. The method is based on the fact that orbital motion 
tends to be important for close binary lenses while it is negligible 
for wide binary lenses. We demonstrate the usefulness of the method 
in the interpretation of an actually observed lensing event 
MOA-2011-BLG-040/OGLE-2011-BLG-0001.

\section{Degeneracy and Resolution}

The existence of an additional lens component can make lensing light 
curves greatly different from that of a single-lens event. The most 
dramatic feature of binary lensing is caustics, which represent 
positions on the source plane where the lensing magnification of a 
point source becomes infinite \citep{schneider86}.  As a result, 
binary-lens light curves resulting from caustic-crossing source 
trajectories exhibit characteristic spikes. Binary lensing caustics 
form a single or multiple sets of closed curves each of which is 
composed of concave curves that meet at cusps. The caustic topology 
is broadly classified into three categories depending on the binary 
separation \citep{erdl93}. For a resonant binary with a normalized 
separation similar to the Einstein radius of the lens $(s\sim1)$, a 
single large caustic with six cusps forms around the center of mass 
of the binary. For a wide binary, there exist two four-cusp caustics 
each of which is located close to each lens components. For a close 
binary, there are three caustics with one four-cusp caustic located 
near the barycenter of the binary while the other two three-cusp 
caustics are located away from the center of mass.

The close/wide degeneracy in the interpretation of a binary-lens 
light curve occurs due to the similarity in shape between the 
central caustic of a close binary and either caustic of a wide 
binary \citep{albrow01}. The binary-lens parameters of the pair 
of degenerate close/wide binaries are related by 
\begin{equation}
q_{\rm w}=q_{\rm c} (1-{q_{\rm c}})^{-2}
\end{equation}
and
\begin{equation}
s_{\rm w}=s_{\rm c}^{-1}(1+{q_{\rm c}})({q_{\rm c}}^2-{q_{\rm c}}+1)^{-1/2}, 
\end{equation}
where $(s_{\rm c},q_{\rm c})$ and $(s_{\rm w},q_{\rm w})$ represent 
the projected separations and the mass ratios of the degenerate sets 
of the close and wide binaries, respectively \citep{albrow02}. In 
the present context, it is important to emphasize that the degeneracy 
becomes especially severe when the normalized separation is either 
substantially smaller $(s\ll1)$ or larger $(s\gg1)$ than the Einstein 
radius \citep{shin12}.

The existence of a binary companion additionally affects lensing 
light curves due to the orbital motion of the binary 
\citep{albrow00, penny11, shin11, skowron11}. The effect of the 
orbital motion is twofold. First, it makes the projected binary 
separation vary in time, causing the shape and size of a caustic 
to vary during the event. Second, it also causes the projected 
binary axis to rotate with respect to the source trajectory, which 
is equivalent to the lens-source relative motion deviating from rectilinear.

% Figure 1 ----------------------------------------------------------------------------------------
\begin{figure}[ht]
\epsscale{1.15}
\plotone{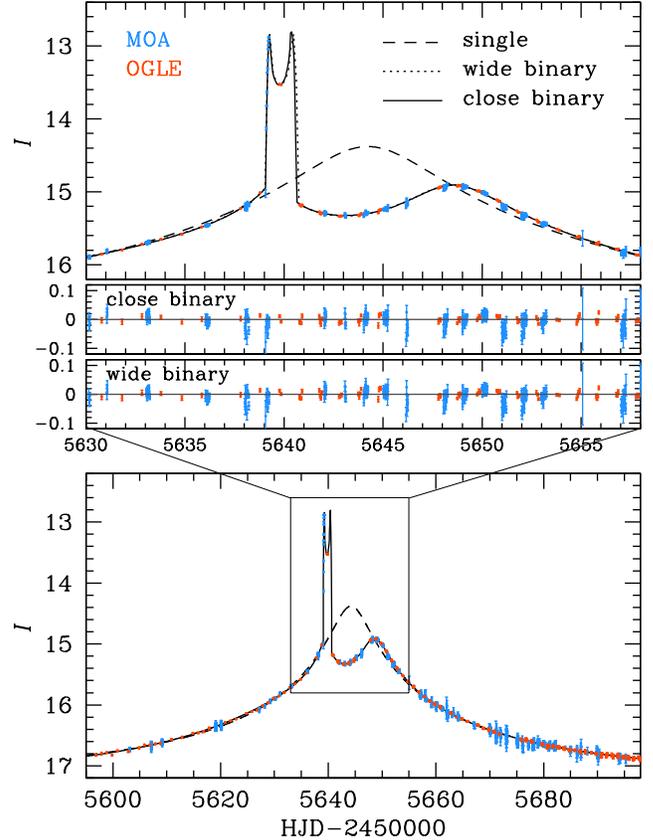}
\caption{\label{fig:one}
Light curve of MOA-2011-BLG-040/OGLE-2011-BLG-0001. Also drawn 
are the best-fit curves based on single-lens, wide-binary, and 
close-binary lens models. We note that the close and wide-binary 
models are too degenerate to be visually distinguished. 
}\end{figure}
% --------------------------------------------------------------------------------------------------

The orbital effect of a binary lens can provide a useful tool 
that allows one to resolve the close/wide degeneracy. The 
physical Einstein radius of a typical Galactic bulge event is 
$\sim$1 -- 2 AU. Then, orbital periods of wide binaries would 
be substantially longer than a year. Considering that typical 
durations of binary-induced deviations are of order of 10 days, 
then, the orbital effect on the light curve of a wide binary 
event would be negligible. By contrast, orbital periods of 
close binaries can be small and thus the change of the lens 
position caused by the orbital motion during the deviating 
part of a lensing event can be important for some close 
binary events. Hence, if it is found that the orbital effect 
is needed to describe an observed light curve, the close 
binary interpretation would be the correct one between two the 
possible close/wide binary interpretations. On the other hand, 
if the orbital motion is constrained to be extremely small, 
this would not absolutely rule out the close-binary solution, 
because the projected separation of a binary with a large
semi-major axis can be small when the binary axis is aligned with
the line of sight. However, because such extreme projections are 
a priori unlikely, strong upper limits on orbital motion would 
statistically favor wide solutions.

% Figure 2 ----------------------------------------------------------------------------------------
\begin{figure}[ht]
\epsscale{1.15}
\plotone{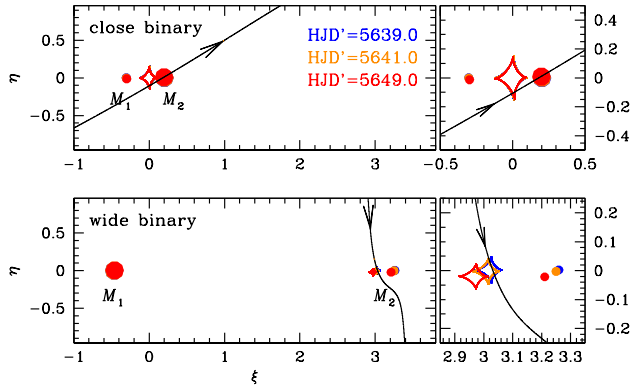}
\caption{\label{fig:two}
Geometry of the lens systems corresponding to the close 
(upper panel) and wide-binary (lower panel) solutions. 
In each panel, the cuspy figures indicate the caustics 
at the times marked in the upper panel. The dots marked 
by $M_1$ and $M_2$ represent the binary components, where 
the bigger dot represents the heavier component. The curve 
with an arrow represents the source trajectory. Right panels 
show enlargements of the region around the caustic where the 
deviation occurred.  The coordinates $(\xi,\eta)$ are centered 
at the center of mass of the binary and all lengths are scaled 
by the angular Einstein radius corresponding to the total mass 
of the binary lens. The $\xi$-axis is set as the binary axis at 
the time ${\rm HJD'}={\rm HJD}-2450000=5640.0$.
}\end{figure}
% --------------------------------------------------------------------------------------------------

\section{Demonstration}

We apply the orbit-based method of resolving the close/wide 
binary degeneracy to an actually observed event 
MOA-2011-BLG-040/OGLE-2011-BLG-0001. The event was discovered 
from the survey conducted by the Microlensing Observations in 
Astrophysics (MOA) group on 2011-Mar-12 toward the Galactic 
bulge at $(\alpha,\delta)_{\rm J2000}=(17^{\rm h}54^{\rm m}
00^{\rm s}\hskip-2pt .02, -29^\circ06'05''\hskip-2pt.70)$, 
i.e., $(l,b)=(0.799^\circ,-1.657^\circ)$, based on observation 
using the 1.8 m telescope at Mt. John observatory in New Zealand 
\citep{sumi11}. The event was also independently discovered in 
May 2011 by the Optical Gravitational Lensing Experiment (OGLE), 
immediately after commissioning of the Early Warning System (EWS) 
of microlensing event detections \citep{udalski03} for OGLE-IV 
phase. OGLE-IV has been conducted since March 2010 with the 1.3 m 
Warsaw telescope and 32 CCD detector mosaic camera at Las Campanas 
Observatory in Chile. Reductions of data were processed 
using photometry codes developed by the individual groups.

Figure \ref{fig:one} shows the light curve of 
MOA-2011-BLG-040/OGLE-2011-BLG-0001. It exhibits a deviation 
from the unperturbed single-lens light curve near the peak. 
The deviation is composed of multiple features. One  
at ${\rm HJD'}(={\rm HJD}-2450000)\sim5639$, shows a typical 
caustic-crossing feature of a strong spike. Another feature is 
the bump at ${\rm HJD'}\sim5649$. Located between the two 
features is an extended negative deviation region that lasted 
for $\sim$6 days.

In modeling the light curve of the event, we search for 
a solution of lensing parameters that best describes the 
observed light curve. For the basic description of a 
binary-lens light curve, seven lensing parameters are 
needed. The first three describe the relative lens-source 
motion, including the time of the closest of the source star 
approach to the center of mass of the binary lens, $t_0$, 
the lens-source separation at that moment, $u_0$, and the 
time scale for the source to cross the Einstein radius, 
$t_{\rm E}$ (Einstein time scale). Another three parameters 
are related to the binary nature of the 
lens, including the normalized projected binary separation, 
$s$, the mass ratio between the binary components, $q$, and 
the angle between the source trajectory and the binary axis, 
$\alpha$. The last parameter is $\rho_{\star}=\theta_{\star}
/\theta_{\rm E}$, the angular source radius $\theta_{\star}$ 
normalized by the angular Einstein radius. The normalized 
source radius $\rho_{\star}$ is needed to account for the 
finite-source effect which affects lensing magnifications during 
caustic crossings or approaches. With these parameters, the 
solution of the lensing parameters is searched by minimizing 
$\chi^2$ in a parameter space. This is done by a combination of 
grid search and down-hill approach. The grid search is done 
in the space of the parameters ($s$,$q$,$\alpha$), which are 
related to features of lensing curves in a complex way. For 
the $\chi^2$ minimization in the down-hill approach, we use 
the Markov Chain Monte Carlo method. Considering the possibility 
of the existence of degenerate close and wide binary solutions, 
the search is done in the parameter space that is wide enough to 
encompass both binary solutions. From this search, we find a pair 
of degenerate close and wide binary solutions.

Although the overall shape of the observed light curve can be 
fitted with the basic binary lensing parameters, there are 
significant residuals to the fit. This gives a hint that 
second-order effects should be taken into account. 
Considering a long duration of the event, which lasted 
$\sim$100 days, we examine the effect of the lens orbital 
motion and the parallax effect. The latter effect is caused 
by the positional change of the observer because of the 
Earth's orbital motion around the Sun \citep{gould92}. It is 
known that the parallax effect results in a long-term deviation 
in lensing light curves similar to the effect of the binary-lens 
orbital motion \citep{batista11} and thus we consider both 
effects simultaneously. To first order approximation, the lens 
orbital motion is described by two parameters $ds/dt$ and 
$d\alpha/dt$, which represent the change rates of the 
normalized binary separation and the source trajectory 
angle, respectively.  The parallax effect requires another 
two parameters $\pi_{{\rm E},N}$ and $\pi_{{\rm E},E}$, 
which are the components of the lens parallax vector 
$\pivec_{\rm E}$ projected on the sky along the north 
and east equatorial coordinates, respectively. The direction 
of the parallax vector corresponds to the relative lens-source 
motion in the frame of the Earth at a specific time of the event. 
The amplitude of the parallax vector is the ratio of the size of 
the Earth's orbit to the Einstein radius projected on the observer 
plane. We find that taking account of these second-order effects 
significantly improves the fit for both the close and wide binary 
solutions.

% Table 1 -------------------------------------------------------------------
\begin{deluxetable}{lrr}
\tablecaption{Model Parameters\label{table:one}}
\tablewidth{0pt}
\tablehead{
%---------------------------------------------------
\multicolumn{1}{c}{Parameter} &
\multicolumn{1}{c}{Close} &
\multicolumn{1}{c}{Wide} 
}
\startdata
% -----------------------------------------------
$\chi^2/{\rm dof}$             &  7734/7955         &  7856/7955         \\ 
$t_0$ (${\rm HJD'}$)           &  5642.40$\pm$0.02  &  5587.30$\pm$0.27  \\
$u_0$                          & -0.0914$\pm$0.0008 &  2.9765$\pm$0.0084 \\
$t_{\rm E}$ (days)             &  53.14$\pm$0.39    &  130.59$\pm$0.44   \\
$s_0$                          &  0.506$\pm$0.002   &  3.718$\pm$0.007   \\
$q$                            &  1.502$\pm$0.013   &  0.142$\pm$0.002   \\
$\alpha_0$                     & -0.530$\pm$0.002   &  1.447$\pm$0.001   \\
$\rho_{\star}$ ($10^{-3}$)     &  2.15$\pm$0.02     &  0.88$\pm$0.01     \\
$\pi_{{\rm E},N}$              & -0.041$\pm$0.067   & -0.065$\pm$0.003   \\
$\pi_{{\rm E},E}$              &  0.127$\pm$0.007   &  0.185$\pm$0.004   \\
$ds/dt$ (${\rm yr}^{-1}$)      & -0.347$\pm$0.027   & -2.040$\pm$0.019   \\
$d\alpha/dt$ (${\rm yr}^{-1}$) &  1.744$\pm$0.074   & -0.273$\pm$0.008
% -----------------------------------------------
\enddata
\tablecomments{ 
${\rm HJD'}={\rm HJD}-2450000$. The lensing parameters $t_0$ and 
$u_0$ are measured with respect to the barycenter of the binary lens.
The parameters $s_0$ and $\alpha_0$ denote the binary separation and the 
source trajectory angle at ${\rm HJD}'=5640.0$, respectively.
}
\end{deluxetable}
% --------------------------------------------------------------------------

% Figure 3 ----------------------------------------------------
\begin{figure*}[ht]
\epsscale{1.15}
\plotone{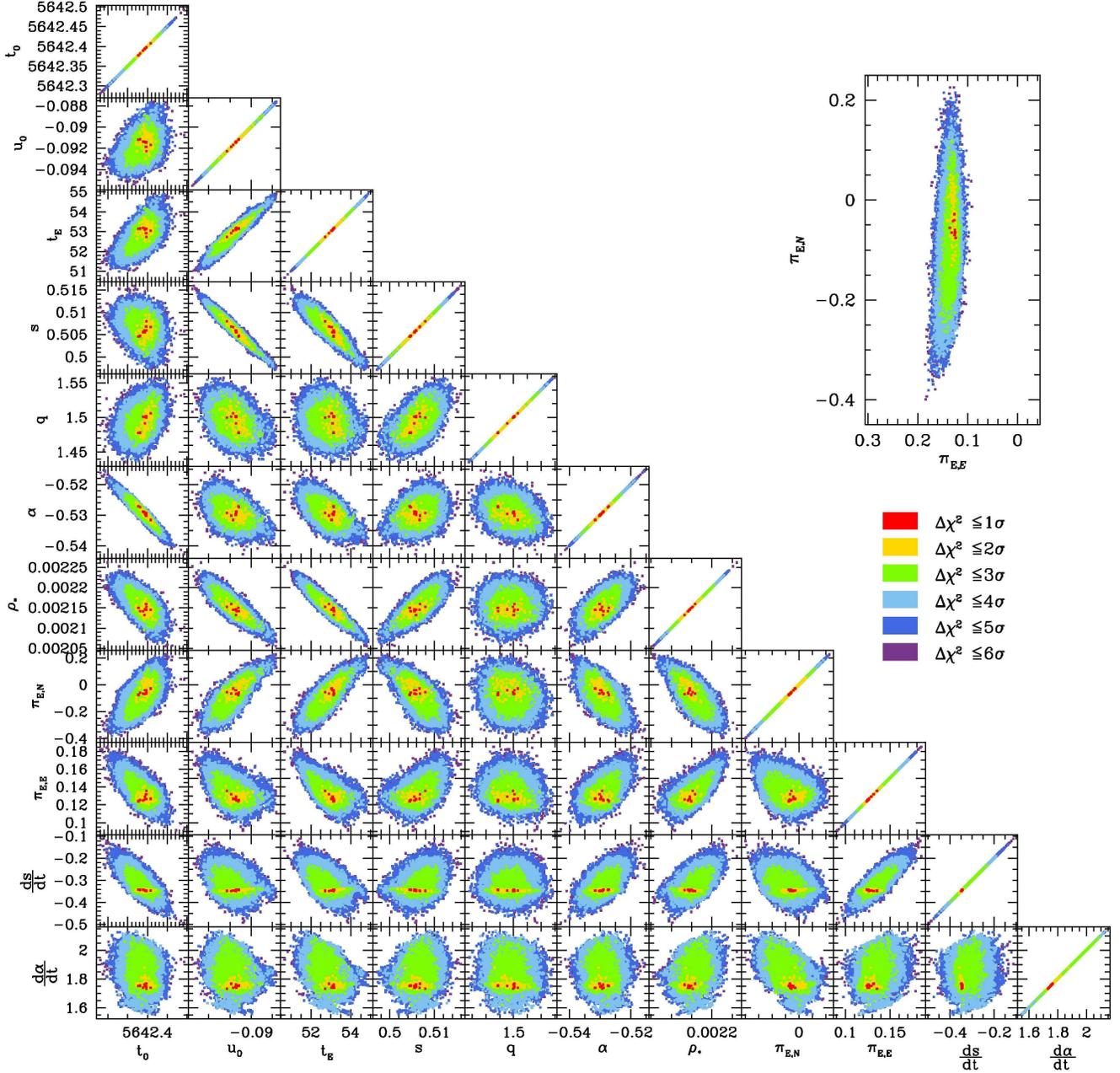}
\caption{\label{fig:three}
Distributions of $\Delta\chi^2$ in the space of lensing parameters. 
For better view, the distribution of the lens-parallax parameters, 
$\pi_{{\rm E},N}$ and $\pi_{{\rm E},E}$, are presented in a separate 
panel on the upper right corner.
}\end{figure*}
% -------------------------------------------------------------

% Figure 4 ----------------------------------------------------
\begin{figure}[ht]
\epsscale{1.15}
\plotone{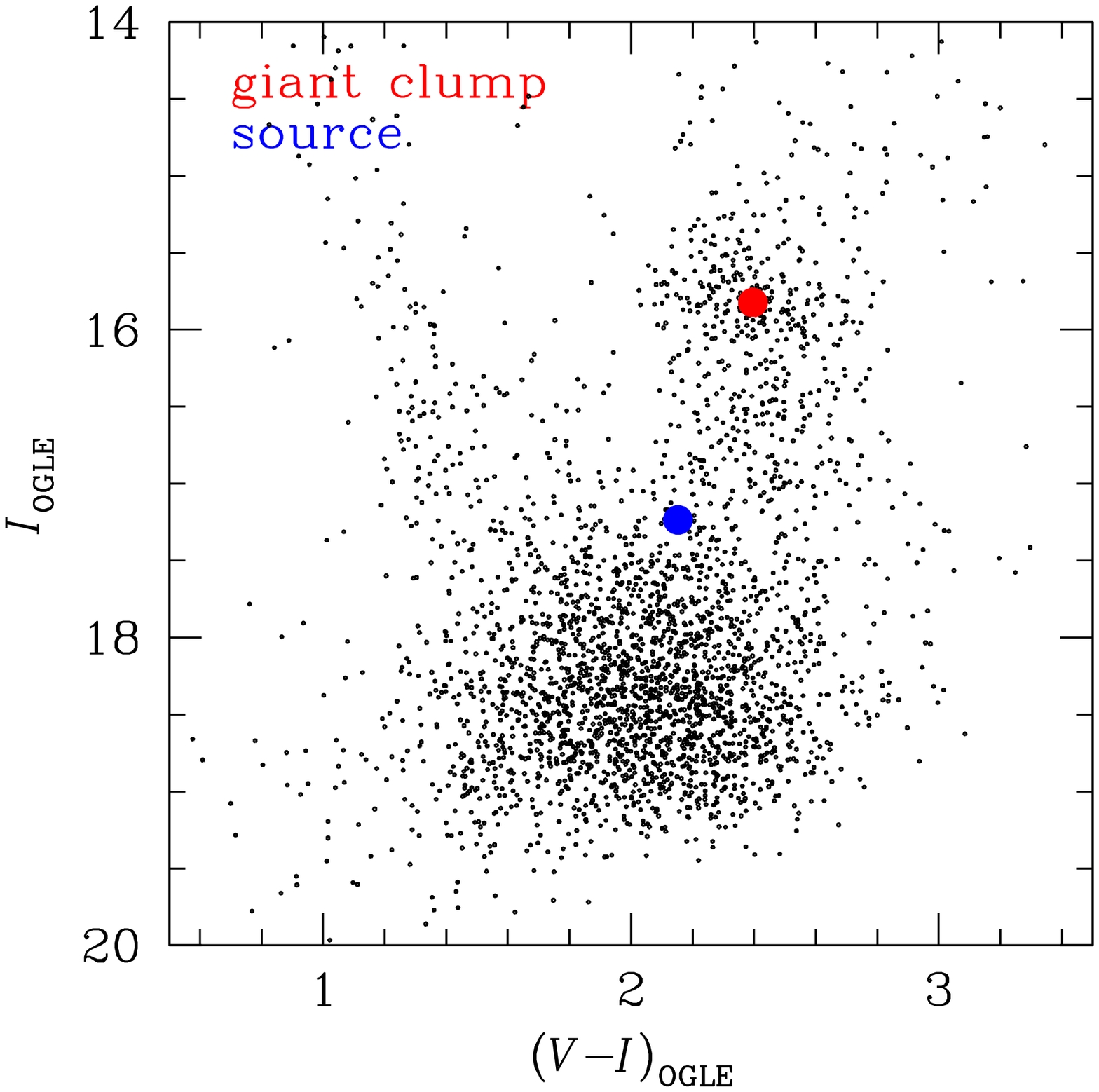}
\caption{\label{fig:four}
OGLE color-magnitude diagram of stars in the field where the 
binary-lens event MOA-2011-BLG-040/OGLE-2011-BLG-0001 occurred. 
The red and blue dots represent the centroid of the giant clump 
and the location of lensed star, respectively
}\end{figure}
% -------------------------------------------------------------

In Figure \ref{fig:one}, we present the model light curves on 
top of the observed data for both the close and wide binary 
solutions. The lensing parameters of the corresponding solutions 
are listed in Table \ref{table:one}. It is found that the close 
binary solution provides a better fit than the wide binary 
solution with $\Delta\chi^2\sim122$. However, the degeneracy is 
quite severe and thus it is difficult to visually distinguish 
the two model curves. To be noted is that the obtained lensing 
parameters $s$ and $q$ of the pair of degenerate solutions well 
follow the relations in equations (1) and (2), suggesting that 
the degeneracy is rooted in the symmetry of the lens-mapping equation. 
In Figure \ref{fig:two}, we also present the geometry of the 
lens system corresponding to the individual solutions. For both 
solutions, the caustic-crossing feature of the light curve occurred 
when the source crossed a tip of a caustic and the bump was produced 
when the source approached a cusp of the caustic. The extended 
negative deviation region occurred when the source passed along 
the fold of the caustic. Figure \ref{fig:three} shows  $\Delta\chi^2$ 
distributions in the lensing-parameter spaces presented to 
show uncertainties and correlations between lensing parameters. 
The presented distributions are for the close-binary solution, 
which provides a better fit.

For both solutions, the orbital effect is important as indicated by 
the large values of the orbital parameters $ds/dt$ and $d\alpha/dt$. 
According to the definition, $ds/dt=1.0$ implies that the binary 
separation changes by $s = 1.0$ per year.  Similarly, $d\alpha/dt=1.0$ 
implies that the binary axis rotates by $1$ radian per year. The 
obtained orbital parameters are $ds/dt=-0.35$ and $d\alpha/dt=1.74$ 
for the close binary model and $ds/dt=-2.04$ and $d\alpha/dt=-0.27$ 
for the wide binary model. Considering that the binary-induced 
deviations lasted $\sim 15$ days, the changes of the binary 
separation and the rotation angle over the course of lensing 
magnification are very significant for both solutions. From 
comparison of the model light curves obtained with and without 
considering the orbital effect, we find that the orbital effect 
is especially important for the description of the extended 
negative perturbation region.

The severity of the orbital effect strongly suggests that the 
event was produced by a close binary rather than a wide binary. 
For a quantitative proof of this, we evaluate the ratio of 
transverse kinetic to potential energy
\begin{equation}
\left({\rm KE \over PE}\right)_\perp
= {{{(r_{\perp}/{\rm AU})}^3}\over{{8\pi^2 (M/M_{\odot})}}} 
\left[ \left({{{1}\over{s}}{{ds}\over{dt}}}\right)^2 + \left( {{d\alpha}\over{dt}} \right)^2 \right],
\end{equation}
which must obey $({\rm KE/PE})_\perp \leq {\rm KE/PE}$, where 
KE/PE is the ratio of (3-dimensional) kinetic to potential 
energy \citep{dong09}. Here $M$ is the total mass of the binary 
system, $r_{\perp}=sD_{\rm L}\theta_{\rm E}$ is the physical 
projected separation between the lens components, and $D_{\rm L}$ 
represents the distance to the lens. An energy ratio greater 
than unity implies that a binary system is dynamically unbound.

To determine the energy ratio, it is required to measure the 
total mass and distance to the lens. We determine these 
quantities from the measured lens parallax and the angular 
Einstein radius by
\begin{equation}
M={\theta_{\rm E}\over{\kappa\pi_{\rm E}}},
\end{equation}
and
\begin{equation}
D_{\rm L}={{\rm AU}\over{\pi_{\rm E}\theta_{\rm E}+\pi_{\rm S}}},
\end{equation}
where $\pi_{\rm E}=({\pi_{{\rm E},N}}^2+{\pi_{{\rm E},E}}^2)^{1/2}$, 
$\pi_{\rm S}={\rm AU}/{D_{\rm S}}$, and $D_{\rm S}$ represents the 
distance to the source star. The Einstein radius is measured by 
$\theta_{\rm E}=\theta_{\star}/\rho_{\star}$.  The normalized 
source radius $\rho_{\star}$ is determined from light curve 
fitting. In this process, we consider the limb-darkening variation 
of the source star's surface by modeling the surface-brightness 
profile as $S_{\lambda}=({\rm F}_{\lambda}/\pi{\theta_{\star}}^2)
[1-\Gamma_{\lambda}(1-3\cos{\psi}/2)]$, where $\Gamma_{\lambda}$ is 
the linear limb-darkening coefficient, ${\rm F}_{\lambda}$ is the 
source star flux, and $\psi$ is the angle between the normal to 
the source star's surface and the line of sight toward the star. 
We adopt $\Gamma_{{\it R}}=0.53$ and $\Gamma_{{\it I}}=0.44$ based 
on the source type determined by the procedure described below. 
We estimate the angular source radius $\theta_{\star}$ 
from the dereddened color $(V-I)_0$ and brightness $I_0$ using 
an instrumental color-magnitude diagram \citep{yoo04}.  For the 
event, we estimate the instrumental color of the source based on 
linear regression method by using multi-band ($V$ and $I$) OGLE 
data. Then, we measure the offset $\Delta[(V-I),I]=
(-0.24,+1.41)$ of the source star from the centroid of bulge 
giant clump, whose dereddened position is known independently, 
$[(V-I),I]_{0,c}=(1.06, 14.41)$ \citep{bensby11, nataf12}. 
This yields the dereddened color and magnitude of the source 
star $(V-I,I)_{0,{\rm S}}=(0.82,15.79)$, indicating that the 
source is a low-luminosity bulge giant. Figure \ref{fig:four} 
shows the locations of the source star and the centroid of 
giant clump in the color-magnitude diagram constructed based 
on OGLE data. Once the dereddened $V-I$ color of the source 
star is measured, it is translated into $V-K$ color by using 
the $V-I$ versus $V-K$ relations of \citet{bessell88} and then 
the angular source radius is estimated by using the relation 
between $V-K$ colors and angular stellar radii given by 
\citet{kervella04}.

In Table \ref{table:two}, we list the projected kinetic to 
potential energy ratios for the close and wide binary solutions 
along with the determined physical quantities. For the close 
binary solution, this ratio [$({\rm KE}/{\rm PE})_\perp=0.4$] is 
smaller than unity, implying that the binary lens system is 
dynamically stable. By contrast, for the wide binary solution, 
this ratio [$({\rm KE}/{\rm PE})_\perp=9.8$] is much larger 
than unity, implying that the binary system is dynamically 
{\it unstable} and thus the wide binary cannot be the solution 
of the observed light curve.  This demonstrates that detections 
of orbital effect can be used to resolve severe binary degeneracies.
By resolving the degeneracy, the physical parameters of the lens are 
uniquely determined. The measured masses of the binary components are 
0.65 $M_{\odot}$ and 0.43 $M_{\odot}$, which correspond to K and M-type 
dwarfs, respectively. The distance to the lens is $\sim$3.5 kpc from the Earth.

% Table 2 ------------------------------------------------------------------
\begin{deluxetable}{lcc}
\tablecaption{Physical lens parameters\label{table:two}}
\tablewidth{0pt}
\tablehead{
% ---------------------------------------------------------------------------
\multicolumn{1}{c}{Parameter} &
\multicolumn{1}{c}{Close} & 
\multicolumn{1}{c}{Wide} 
}
\startdata
% ---------------------------------------------------------------------------
$M$                 ($M_{\odot}$)         & 1.08$_{-0.24}^{+0.05}$  & 1.80$\pm$0.05 \\ 
$M_{\rm 1}$         ($M_{\odot}$)         & 0.43$_{-0.10}^{+0.02}$  & 1.58$\pm$0.04 \\
$M_{\rm 2}$         ($M_{\odot}$)         & 0.65$_{-0.14}^{+0.03}$  & 0.22$\pm$0.01 \\
$\theta_{\rm E}$    (mas)                 & 1.17$\pm$0.01           & 2.88$\pm$0.02 \\
$\mu$               (mas ${\rm yr}^{-1}$) & 8.04$\pm$0.02           & 8.05$\pm$0.06 \\
$D_{\rm L}$         (kpc)                 & 3.54$_{-0.45}^{+0.06}$  & 1.45$\pm$0.03 \\
$({\rm KE/PE})_\perp$                     & 0.38$_{-0.05}^{+0.02}$  & 9.84$\pm$0.58
% ---------------------------------------------------------------------------
\enddata
\tablecomments{
$M$: total mass of the close binary, $M_{\rm 1}$ and $M_{\rm 2}$: masses of 
the binary components, $\theta_{\rm E}$: angular Einstein radius, $\mu$: 
relative lens-source proper motion, $D_{\rm L}$: distance to the lens, 
$({\rm KE/PE})_\perp$: transverse kinetic to potential energy ratio.
}
\end{deluxetable}
% ---------------------------------------------------------------------------

\section{Summary and Conclusion}

We proposed a method of resolving the close/wide degeneracy in microlensing 
light curves using the effect of the lens orbital motion on light curves. 
The method is based on the fact that the orbital effect tends to be important 
for close binaries while it is negligible for wide binaries. We demonstrated 
the usefulness of the method by applying it to an actually observed lensing 
event MOA-2011-BLG-040/OGLE-2011-BLG-0001.  From this, we were able to 
uniquely identify that the lens is composed of two K and M-type dwarfs located 
at $\sim$3.5 kpc from the Earth.  Considering that the orbital effect becomes 
more important with the decrease of the binary separation, the proposed 
method would be important especially in resolving binaries with very close 
and wide separations for which the degeneracy is very severe.

\mbox{}

\acknowledgments 
Work by C.\ Han was supported by the Creative Research Initiative Program 
(2009-0081561) of National Research Foundation of Korea.
The OGLE project has received funding from the European Research Council 
under the European Community's Seventh Framework Programme (FP7/2007-2013) 
/ ERC grant agreement no. 246678.  
The MOA experiment was supported by grants JSPS22403003 and JSPS23340064.
T.S. was supported by the grant JSPS23340044.
Y. Muraki acknowledges support from JSPS grants JSPS23540339 and JSPS19340058.
A. Gould acknowledges support from NSF AST-1103471 and NASA grant NNG04GL51G.


\begin{thebibliography}{99}

\bibitem[Albrow et al.(2000)]{albrow00}
Albrow, M. D., Beaulieu, J.-P., Caldwell, J. A. R., et al.\ 2000, \apj, 534, 894

\bibitem[Albrow et al.(2001)]{albrow01}
Albrow, M. D., An, J., Beaulieu, J.-P., et al.\ 2001 \apj, 549, 759

\bibitem[Albrow et al.(2002)]{albrow02}
Albrow, M. D., An, J., Beaulieu, J.-P., et al.\ 2002, \apj, 572, 1031

\bibitem[An(2005)]{an05}
An, J. H. 2005, \mnras, 356, 1409

\bibitem[Batista et al.(2011)]{batista11}
Batista, V., Gould, A., Dieters, S., et al.\ 2011, \aap, 529, 102

\bibitem[Bensby et al.(2011)]{bensby11}
Bensby, T., Ad\'en, D., Mel\'endez, J., et al.\ 2011, \aap, 533, 134

\bibitem[Bessell \& Brett(1988)]{bessell88}
Bessell, M. S., \& Brett, J. M.\ 1988, PASP, 100, 1134

\bibitem[Bond et al.(2001)]{bond01}
Bond, I. A., Abe, F., Dodd, R. J., et al.\ 2001, \mnras, 32, D.7, 868

\bibitem[Dominik(1999)]{dominik99}
Dominik, M.\ 1999, \aap, 349, 108

\bibitem[Dong et al.(2009)]{dong09}
Dong, S., Gould, A., Udalski, A., et al.\ 2009 \apj, 695, 970

\bibitem[Erdl \& Schneider(1993)]{erdl93}
Erdl, H., \& Schneider, P.\ 1993, \aap, 268, 453

\bibitem[Gould(1992)]{gould92}
Gould, A.\ 1992, \apj, 392, 442

\bibitem[Griest \& Safazadeh(1998)]{griest98}
Griest, K., \& Safazadeh, N.\ 1998, \apj, 500, 37

\bibitem[Kervella et al.(2004)]{kervella04}
Kervella, P., Th\'evenin, F., Di Folco, E., \& S\'egransan, D.\ 2004, \aap, 426, 297

\bibitem[Mao \& Paczy\'nski(1991)]{mao91}
Mao, S., \& Paczy\'nski, B.\ 1991, \apj, 374, L37

\bibitem[Nataf et al.(2012)]{nataf12}
Nataf, D. M., Gould, A., Fouqu\'e, P., et al.\ 2012, \apj, submiited

\bibitem[Penny et al.(2011)]{penny11}
Penny, M. T., Mao, S., \& Kerins, E.\ 2011, \mnras, 412, 607

\bibitem[Schneider \& Wei\ss(1986)]{schneider86}
Schneider, P., \& Wei\ss, A.\ 1986, \aap, 164, 237

\bibitem[Shin et al.(2011)]{shin11}
Shin, I.-G., Udalski, A., Han, C., et al.\ 2011, \apj, 735, 85 

\bibitem[Shin et al.(2012)]{shin12}
Shin, I.-G., Choi, J.-Y., Park, S.-Y., et al.\ 2012, \apj, 746, 127

\bibitem[Skowron et al.(2011)]{skowron11}
Skowron, J., Udalski, A., Gould, A., et al.\ 2011, \apj, 738, 87

\bibitem[Sumi et al.(2003)]{sumi03}
Sumi, T., Abe, F., Bond, I. A., et al.\ 2003, \apj, 591, 204

\bibitem[Sumi et al.(2011)]{sumi11}
Sumi, T., Kamiya, K., Bennett, D. P., et al.\ 2011, Nature, 473, 349

\bibitem[Udalski(2003)]{udalski03}
Udalski, A.\ 2003, Acta Astron., 53, 291

\bibitem[Yoo et al.(2004)]{yoo04}
Yoo, J., DePoy, D. L., Gal-Yam, A., et al.\ 2004, \apj, 603, 139

\end{thebibliography}
\end{document}